\documentclass[aip,jap,reprint]{revtex4-1}

\usepackage[utf8]{inputenc}
\usepackage{amsmath,amssymb,amsfonts}
\usepackage{graphicx,epsfig}

\usepackage[squaren]{SIunits}

\providecommand{\vect}[1]{\boldsymbol{#1}}

\begin{document}

\title{Real-Space Berry Phases -- Skyrmion Soccer}

\author{Karin Everschor-Sitte}
\affiliation{The University of Texas at Austin, Department of Physics, 2515 Speedway, Austin, TX 78712, USA}
\author{Matthias Sitte}
\affiliation{The University of Texas at Austin, Department of Physics, 2515 Speedway, Austin, TX 78712, USA}

\begin{abstract}
Berry phases occur when a system adiabatically evolves along a closed
curve in parameter space. This tutorial-like article focuses on Berry
phases accumulated in real space. In particular, we consider the situation
where an electron traverses a smooth magnetic structure, while its
magnetic moment adjusts to the local magnetization direction. Mapping the
adiabatic physics to an effective problem in terms of emergent fields
reveals that certain magnetic textures -- skyrmions -- are tailormade to
study these Berry phase effects.
\end{abstract}

\date{\today}

\maketitle

Over the last decades physicists have realized that a seemingly harmless
phase -- the Berry phase --  leads to a lot of interesting physical
consequences like quantum,\cite{Thouless1982}
anomalous,\cite{Nagaosa2010}
topological,\cite{Bruno2004, *Onoda2004, Neubauer2009} and
spin\cite{Murakami2003, *Sinova2004, *Culcer2004} Hall effects,
electric polarization,\cite{King1993, *Resta1994}
orbital magnetism,\cite{Thonhauser2005, *Xiao2005}
quantum charge pumping,\cite{Thouless1983} etc.
(For a review of Berry phase effects on electronic properties see
Ref.~\citenum{Xiao2010}.)
Combining the Berry phase concepts with the physics of an electric
current traversing a peculiar whirl-like magnetic texture first
discovered\cite{Muehlbauer2009} in 2009 -- magnetic skyrmions --
reveals the elegance of these phases in a vivid way.

This paper is structured as follows: First, we review the general concept
of Berry phases and introduce the main formulas. In
Sec.~\ref{sec:realspaceberryphases}, we focus on real-space Berry phase
effects. Here we discuss the example of a free electron traversing a
spatially or temporally inhomogeneous, smooth magnetic structure. In the
adiabatic limit, the magnetic moment of the electron adjusts constantly to
the local magnetization direction and thereby the electron picks up a
Berry phase. To interpret the physical consequences of these Berry phases,
we review the mapping onto a problem, where the electron moves in a
uniform Zeeman magnetic field, but instead ``feels'' an emergent electric
field and an emergent orbital magnetic field, leading, for example, to the
topological Hall effect\cite{Neubauer2009}. In Sec.~\ref{sec:skyrmions}, we first introduce
skyrmions in a formal way and then later focussing on skyrmions in
magnetic systems, in particular those in a certain material class denoted
as B20 materials. We discuss that those magnetic textures are tailormade
to observe the emergent electric fields introduced in
Sec.~\ref{subsec:emergentfields}. In Sec.~\ref{ssec:SkyrmionSpintronics} we
switch the perspective and discuss the back reaction of the electronic
Berry phase effects on the skyrmion lattice.

\section{Berry phases}
\label{sec:berryphases}

\begin{figure}[tb]
\centering
\includegraphics[width=0.18\textwidth]{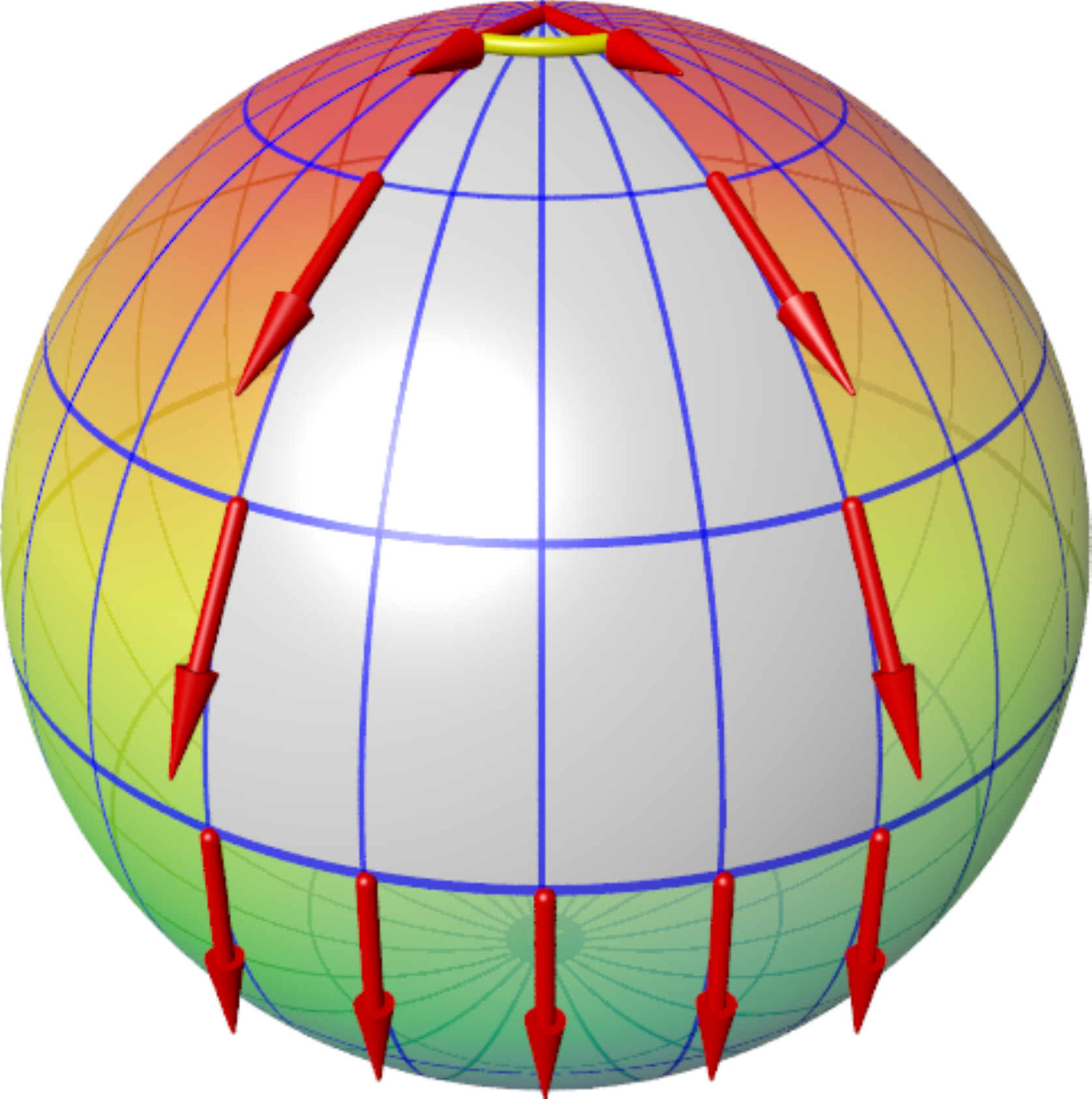}
\caption{%
(Color online) Parallel transport of a vector along a closed path on the
sphere $S_{2}$ leads to a geometric phase between initial and final state.
}
\label{fig:ParallelTransport}
\end{figure}

Berry phases occur in all parts of physics and arise whenever a system
adiabatically evolves along a cyclic process $\mathcal C$ in some
parameter space $\vect X$. A simple classical example is the parallel
transport of a vector along a closed curve on a sphere (see
Fig.~\ref{fig:ParallelTransport}). Although the system returns to its
initial state it aquires a geometric phase characterized by the enclosed
area on the sphere (Gauss-Bonnet theorem).

A quantum-mechanical example is a system whose states are non-degenerate
and which evolves adiabatically in the parameter space $\vect{X} = \vect{X}(t)$.
In that case, the solution of the time-dependent Schr{\"o}dinger equation
$i \hbar \partial_t \left | \psi(t) \right \rangle = H(\vect X(t)) \left| \psi(t) \right \rangle$
is given by\cite{Berry1984}
\begin{equation}
\left | \psi(t) \right \rangle = \sum_n a_n(t_0) e^{i \gamma_n(\mathcal C)}
e^{- \frac{i}{\hbar}
\int_0^t dt' \epsilon_n(\vect X (t'))} \left| \psi_n(t) \right \rangle,
\end{equation}
where $\left| \psi_n(t) \right \rangle$ and $\epsilon_n(\vect X(t))$ are the
time-evolved eigenstates and energies of the Hamiltonian $H(\vect X (t))$,
$\left | \psi(t_0) \right \rangle = \sum_n a_n(t_0) \left| \psi_n(t_0) \right \rangle$
defines the initial state, and $\gamma_n(\mathcal C)$ is the famous Berry phase.

Formally, the Berry phase $\gamma_n(\mathcal C)$ is given by the line
integral of the so-called Berry connection $\vect{\mathcal{A}}_n(\vect X)$
along the closed path $\mathcal{C}$:
\begin{align}
\label{eq:Berryphase}
\gamma_n(\mathcal C) &= \oint_{\mathcal{C}} \vect{\mathcal{A}}_n (\vect X) \cdot d\vect X,\\
\vect{\mathcal{A}}_n(\vect X)&= i \left \langle \psi_n \right | \nabla_{\vect X} \left| \psi_n \right \rangle.
\label{eq:Berryvectorpotential}
\end{align}
Since the latter is a gauge-dependent vector-valued quantity it is also
referred to as the Berry vector potential, in analogy to electrodynamics.
The Berry phase, however, as a closed contour integral is independent of
the gauge choice up to integer multiples of $2\pi$. By means of Stokes
theorem one can reformulate the Berry phase in terms of a surface integral.
For a three dimensional parameter space $\vect X$ this defines the
so-called Berry curvature
\begin{equation}
\vect \Omega_n(\vect X) = \nabla_{\vect X} \times \vect{\mathcal{A}}_n (\vect X)
\label{eqn:berrycurvature}
\end{equation}
which takes a similar role as the magnetic field in electrodynamics (see
below).
The generalization of the Berry curvature to an arbitrary-dimensional
parameter space is given by the antisymmetric rank-2 Berry
curvature tensor
$\Omega_{\mu \nu} (\vect X) = \partial_{X^{\mu}}\mathcal{A}_{\nu}^n - \partial_{X^{\nu}} \mathcal{A}_{\mu}^n$,
where $\Omega_{\mu \nu} = \epsilon_{\mu \nu \xi} (\vect \Omega_n)_{\xi}$.

So far we have not specified the parameter space $\vect X$. In general,
position and momentum degrees of freedom are coupled and therefore
real-space, momentum-space, and even mixed Berry phases occur
simultaneously.\cite{Xiao2010, Freimuth2013} In this paper, we focus on the 
limit, where mainly effects arising from Berry phases collected in 
real-space are important.

\section{Real-space Berry phases}
\label{sec:realspaceberryphases}

An archetypical example of real-space Berry phase physics is a free
electron traversing a spatially or temporally inhomogeneous, smooth
magnetic texture $\vect{M}(\vect{r}, t)$ with constant amplitude
$M = |\vect{M}(\vect{r}, t)|$:
\begin{equation}
\label{eq:schroedinger}
i \hbar \partial_t |\psi\rangle = \left[ \frac{\vect p^2}{2 m} \openone
- J \vect \mu \cdot \vect M ( \vect r, t) \right] |\psi\rangle.
\end{equation}
Here, $\psi = (\psi_{\uparrow}, \psi_{\downarrow})^T$ is the two-component
wave function, $\openone$ is the $2\times2$ unit matrix, $J>0$ is the
strength of the ferromagnetic exchange coupling, and $\vect{\mu}$ is the
magnetic moment of the electron. Note that for a particle with mass $m$,
charge $q$, and Land\'{e} factor $g$ the magnetic moment $\vect{\mu}$ and
the spin $\vect{s}$ are related by $\vect{\mu} = -(g q)/(2m) \vect{s}$. In
particular, for a free electron with charge $q=-e$ the magnetic moment and
the spin are antiparallel:
\begin{equation}
\vect{\mu} = - \frac{g e}{2 m} \vect{s} = - \frac{g \mu_B}{2} \vect{\sigma},
\end{equation}
where $\mu_B$ is the Bohr magneton, and
$\vect{\sigma} = (\sigma_x, \sigma_y, \sigma_z)^T$ denotes the vector of
Pauli matrices. In a ferromagnetic material, the finite magnetization
allows to define majority (minority) spins whose magnetic moments are
parallel (antiparallel) to the magnetization direction. Historically, in
the magnetic materials community the majority and minority spins are
referred to as ``spin-up'' and ``spin-down'' states, while in the
semiconductor community ``spin-up'' and ``spin-down'' refers to the actual
orientation of the electron spin. Here, we follow the latter nomenclature.

In the adiabatic limit, the magnetic moment of the electron adapts
constantly according to the local magnetization direction, and thereby the
electron picks up a Berry phase.\cite{Berry1984, Ye1999, Taguchi2001,
Bruno2004, *Onoda2004, Tatara2007, Binz2008} From this point of view, the
physical consequences of the real-space Berry phase are ``hidden'' in the
spatially and temporally varying spin states which enter the Berry vector
potential [see Eq.~\eqref {eq:Berryvectorpotential}],
$\vect{\mathcal{A}}_{\sigma } = i \langle \psi_{\sigma} | \nabla | \psi_{\sigma} \rangle$
with $\sigma =\ \uparrow, \downarrow$ representing minority and majority
spins, respectively. Note that minority and majority spins acquire opposite
Berry phases when traversing a magnetic texture.

\subsection{Alternative picture: emergent fields}
\label{subsec:emergentfields}

An intuitive way to understand this adiabatic Berry phase physics is to
consider a mapping onto a problem, where the electron moves in a uniform
Zeeman magnetic field, but instead ``feels'' an additional emergent
electric field $\vect{E}^e$ and an additional emergent (orbital) magnetic
field $\vect{B}^e$:\cite{Volovik1987, *Barnes2007, *Yang2009, Ye1999, Bruno2004, *Onoda2004,
 Tatara2007, Binz2008, Zhang2009, Zang2011}
\begin{subequations}
\label{eq:emergentBandEfield}
\begin{align}
\label{eq:emergentBfield}
\vect B^e_i &= \frac{\hbar}{2} \epsilon_{ijk} \hat{\vect M} \cdot (\partial_j
\hat{\vect M} \times \partial_k \hat{\vect M}),\\
\label{eq:emergentEfield}
\vect E^e_i &= \hbar\, \hat{\vect M} \cdot (\partial_i \hat{\vect M} \times
\partial_t \hat{\vect M}),
\end{align}
\end{subequations}
where $\partial_i = \partial/\partial r_i$, and
$\hat{\vect{M}} (\vect{r}, t) = \vect{M}(\vect{r}, t)/M$ is the local
magnetization direction. It is important to note that these emergent
fields lead to Lorentz forces on the electrons that can be measured in
experiments. For example, the emergent magnetic field leads directly to a
contribution to the Hall signal. 
Since the profile of the magnetization
configuration enters in $\vect{B}^e$, this type of Hall signal can be 
denoted as a geometrical Hall effect. In cases, where the magnetic texture
is topologically nontrivial, it is commonly referred to the topological 
Hall effect.
Note that the emergent electric field $\vect{E}^e$
is only finite for a non-collinear, time-dependent magnetic texture. In
Sec.~\ref{ssec:EmergentElectricField}, we discuss the situation of a moving
magnetic texture and its consequences for the corresponding Hall signal.

In the following, we derive the above emergent fields following
Ref.~\citenum{Zhang2009} and show that the emergent magnetic field is indeed
the real-space Berry curvature.\cite{FOOTNOTE1}
The idea is to perform a \emph{local} transformation
$\psi = U(\vect{r}, t) \zeta$, so that the second part of the Hamiltonian
becomes trivial:
\begin{equation}
 - J \vect \mu \cdot \vect M ( \vect r, t) =
\tilde{J} \vect \sigma \cdot \hat {\vect M} ( \vect r, t) \xrightarrow{U(\vect r, t)}
\tilde{J} \sigma_z,
\end{equation}
with $\tilde{J}= J g \mu_B (\hbar/2) M >0$. Hence, one has to rotate the
quantization axis from the $\hat{\vect z}$-direction to the local
magnetization direction $\hat{\vect{M}}(\vect{r}, t)$. Such a unitary
transformation $U$ is given by
\begin{equation}
U= e^{-i (\theta/2) \vect \sigma \cdot \vect n} = \cos(\theta/2) \openone
- i \sin(\theta/2) \vect \sigma \cdot \vect n,
\end{equation}
where $\theta = \arccos(\hat{\vect M} \cdot \hat{\vect z})$ is the angle of
rotation, and
$\vect n = \hat{\vect z} \times \hat{ \vect M}/ |\hat{\vect z} \times \hat{ \vect M}|$
is the rotation axis. Note that we have dropped the explicit space and time
dependence to simplify the notation.  Multiplying Eq.~\eqref{eq:schroedinger}
by $U^{\dagger}$ from the left and inserting $\psi = U \zeta$ leads to a
Schr\"{o}dinger equation for $\zeta$ of the following form:
\begin{equation}
i \hbar \partial_t \zeta=\left[ q^e V^e + \frac{(\vect p \openone - q^e \vect A^e)^2}{2m} +\tilde{J} \sigma_z\right]\zeta.
\label{eq:schroedingerforzeta}
\end{equation}
By analogy to the Hamiltonian of a free electron under the influence of an
electric and orbital magnetic field one can denote the $2 \times 2$ matrices
\begin{subequations}
\label{eq:emergentpotentials}
\begin{align}
V^e &= -(i \hbar/q^e) U^{\dagger} \partial_t U,\\
\vect A^e &= (i \hbar/q^e) U^{\dagger} \nabla U
\end{align}
\end{subequations}
as the emergent ``scalar'' and ``vector'' potentials. At this level the
emergent charge $q^e$ is introduced artifically and actually drops out of
Eq.~\eqref{eq:schroedingerforzeta}. Furthermore, note that this
transformation is exact.

For a magnetization texture $\vect M(\vect r, t)$ which varies smoothly in
space and time, one can treat the scalar and vector potentials $V^e$ and
$\vect{A}^e$ as a perturbation to the unperturbed Hamiltonian
$H_0= \vect p^2/(2m) \openone +\tilde{J} \sigma_z$. In the adiabatic
approximation, $V^e$ and $\vect A^e$ act on each band separately, allowing
us to introduce electromagnetic potentials for both bands:
\begin{subequations}
\begin{align}
 \vect {\mathcal{A}}^e_{\sigma} &= \langle \sigma | \vect A^e | \sigma \rangle =
 (i \hbar/q^e) \langle \psi_\sigma | \nabla | \psi_\sigma \rangle,\\
 \mathcal{V}^e_{\sigma} &= \langle \sigma | V^e | \sigma \rangle =
 -(i \hbar/q^e) \langle \psi_\sigma | \partial_t | \psi_\sigma \rangle,
\end{align}
\end{subequations}
with $\sigma =\ \uparrow, \downarrow$ for minority and majority spins,
respectively, and $|\psi_{\sigma}\rangle = U|\sigma\rangle$. Above
potentials have the same form of the Berry vector potential introduced in
Eq.~\eqref{eq:Berryvectorpotential}.

Finally, introducing for each band an emergent electric field,
$\vect E_{\sigma}^e = -\nabla \, \mathcal{V}^e_{\sigma} - \partial_t \vect{\mathcal{A}}^e_{\sigma}$,
and an emergent magnetic field,
$\vect B_{\sigma}^e = \nabla \times \vect{\mathcal{A}}^e_{\sigma}$,
that are ``felt'' by the electron, it becomes clear that the real-space
Berry curvature acts like an emergent magnetic field, while the mixed
space-time Berry curvature acts like an emergent electric field. By an
explicit calculation one finds
\begin{subequations}
\begin{align}
(\vect B_{\sigma}^e)_i &= \mp \frac{\hbar}{2 q^e}\, \frac{\epsilon_{ijk}}{2} \hat{\vect M} \cdot \bigl(\partial_j \hat{\vect M} \times \partial_k \hat{\vect M}\bigr),\\
(\vect E_{\sigma}^e)_i &= \mp \frac{\hbar}{2 q^e}  \hat{\vect M} \cdot \bigl (\partial_i \hat{\vect M} \times \partial_t \hat{\vect M} \bigr),
\end{align}
\end{subequations}
where the upper (lower) sign corresponds to the band for electrons with
minority (majority) spin. Let us now assign different emergent charges to
the two bands, because the sign of the Berry phase depends on the
orientation of the spin.\cite{Schulz2012} For a minority (majority) spin we
define\cite{FOOTNOTE2} $q^e_\uparrow=-1/2$ ($q^e_\downarrow=1/2$),
leading to the emergent fields given in Eq.~\eqref{eq:emergentBandEfield}.

In principle, any magnetic structure that varies smoothly in position space
leads to an emergent magnetic field [Eq.~\eqref{eq:emergentBfield}]
and thus to a geometrical Hall effect. As it turns out, there are ideal
magnetic textures -- skyrmions -- which are tailored to investigate
these emergent fields.

\section{Skyrmions}
\label{sec:skyrmions}

\begin{figure}[tb]
\centering
\includegraphics[width=0.43\textwidth]{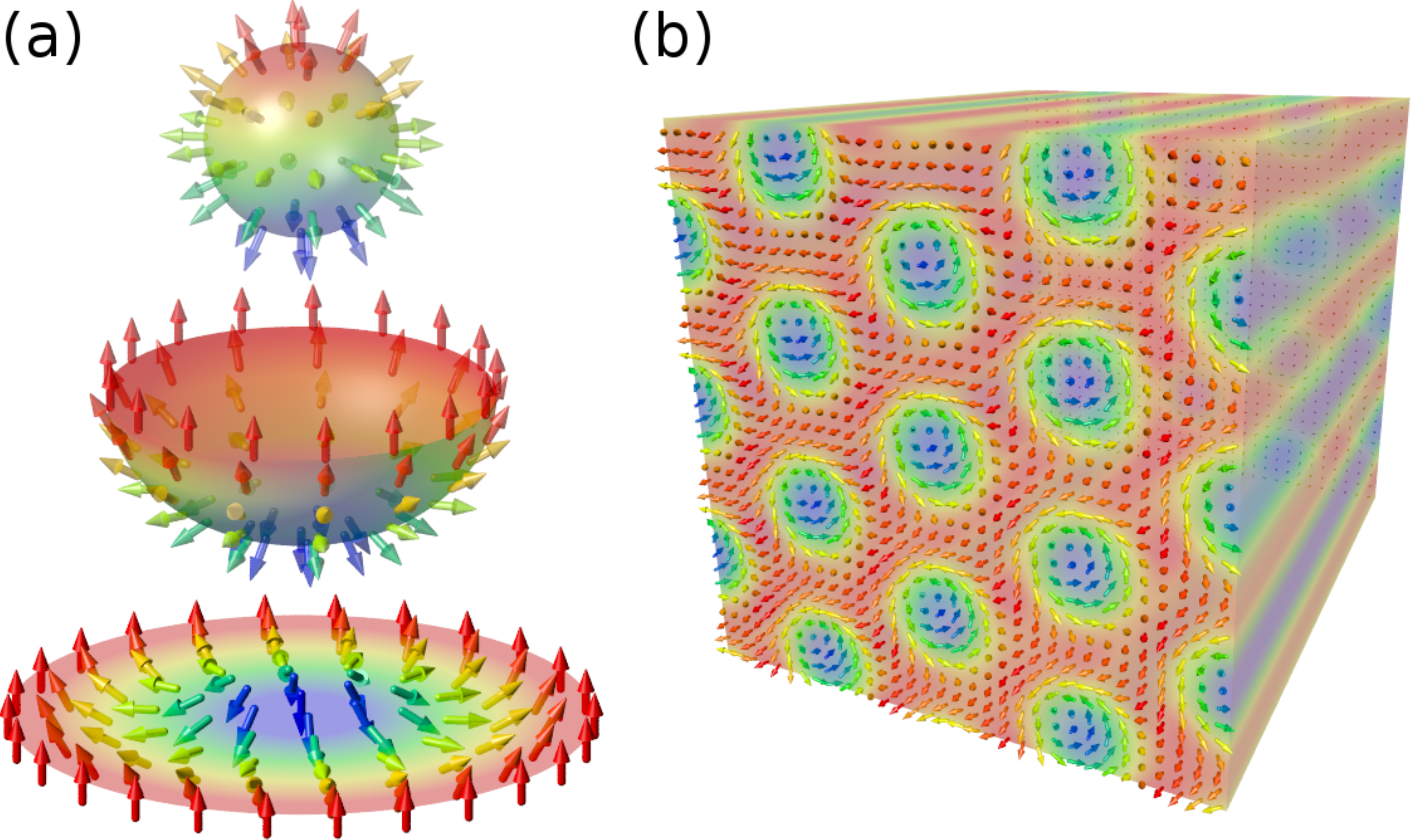}
\caption{%
(Color online)
\textbf{(a)} A (non-chiral) skyrmion configuration (bottom) is obtained by
``unfolding'' a hedgehog (top). Arrows correspond to the local magnetization
direction $\hat{\vect M}= \vect M / |M|$. The color code is chosen according
to the out-of-plane component of the arrows: from red (``up'') over green
(in-plane) to blue (``down'').
\textbf{(b)} Schematic plot of a chiral skyrmion lattice with an additional
in-plane winding of the magnetization.
}
\label{fig:Skyrmions}
\end{figure}

Like the concept of the Berry phase, a skyrmion is a certain mathematical
object which is realized in many areas of physics ranging from nuclear and
particle physics over high-energy physics to condensed matter physics. It
is named after the nuclear physicist Tony Skyrme who in the early 1960's
studied a certain nonlinear field theory for interacting pions, showing
that quantized and topologically stable field configurations -- nowadays
called skyrmions -- do occur as solutions of such field
theories.\cite{Skyrme1961, *Skyrme1962} In the original work, Skyrme
considered three-dimensional versions of skyrmions, but later the notion
of a skyrmion was generalized to arbitrary dimensions: One can define a
skyrmion as a topologically stable, smooth field configuration describing
a non-trivial surjective mapping from coordinate space to an order
parameter space with a non-trivial topology. In the following, we restrict
the discussion to the two-dimensional unit sphere, $S_2$, describing the
magnetization direction $\hat{\vect{M}}$. An intuive picture of a skyrmion
and the mapping to a sphere is shown in Fig.~\ref{fig:Skyrmions}~(a),
where ``infinity'' (the boundary of the skyrmion) is mapped onto the
north pole. Note that a skyrmion is everywhere non-singular and finite.
In contrast to vortices, skyrmions are trivial at infinity, \textit{i.e.},
all arrows at the boundary point in the same direction out of plane.

At the end of the 1980's skyrmion structures were shown to be the
mean-field ground states of certain models for anisotropic,
non-centrosymmetric magnetic materials with chiral spin-orbit interactions
subjected to a magnetic field.\cite{Bogdanov1989, *Bogdanov1994,
*Bogdanov1995} Although some further theoretical works appeared on
magnetic skyrmions and similar textures,\cite{Binz2006a, *Binz2006b,
*Roessler2006, *Fischer2008, *Yi2009, *Han2010, Binz2008} the real
breakthrough was in 2009 when a hexagonal lattice of skyrmion-tubes
perpendicular to a finite, external magnetic field [as sketched in
Fig.~\ref{fig:Skyrmions}~(b)] was experimentally discovered in the cubic
helimagnet manganese silicide (MnSi).\cite{Muehlbauer2009} Since 2009
skyrmions have been observed in many other B20 compounds\cite{Muenzer2009, *Shibata2013,
Neubauer2009, Pfleiderer2010}, including metals, semiconductors, and also
an insulating, multiferroic material.\cite{Seki2012, *Adams2012} The
skyrmion lattice was also confirmed to exist in thin films by a direct
imaging of the real-space magnetic texture by Lorentz transmission
electron microscopy.\cite{Yu2010, Yu2011, *Yu2012a} Moreover, skyrmion textures have
been discussed in thin films in the form of magnetic bubble
domains.\cite{Malozemoff1979, *Yu2012b} Furthermore, skyrmions have been
found on surfaces as a spontaneous magnetic ground state forming a lattice
on the atomic scale.\cite{Heinze2011}

In skyrmion lattices, different length scales 
appear:\cite{Muehlbauer2009,Nagaosa2012} the size of the atomic unit cell 
corresponding to the wavelength of the electrons, the
diameter of the skyrmions, the (non-spinflip) mean-free path, and the much
larger spinflip scattering length. In the adiabatic limit, where the size
of the skyrmions is much larger than the non-spinflip scattering length,
band structure effects are negligible, and one can concentrate on
real-space Berry phases.\cite{Nagaosa2012} In the opposite limit, band 
structure effects are expected to dominate, and momentum-space Berry 
phases become relevant.\cite{Nagaosa2012}

In B20 structures like MnSi a hierachy of energy scales exists which also
determines the size of the skyrmion structures: ferromagnetic exchange
coupling, Dzyaloshinskii-Moriya interaction, and crystalline field
interactions. Since the ratio of the Dzyaloshinskii-Moriya and the
ferromagnetic exchange interaction is small in B20 structures, the
diameter of a skyrmion is much larger than the crystallographic unit cell,
and the skyrmion textures are smooth. Consequently, those magnetic
structures are good candidates to study the physics of real-space Berry
phases.

\subsection{Winding number and emergent magnetic field}

Skyrmions can be classified according to their integer winding number
$\mathcal{W}$ counting the number of times the field configuration wraps
around the whole sphere. To obtain the winding number of a magnetization
configuration which varies in the $xy$ plane as the example shown in the
bottom panel of Fig.~\ref{fig:Skyrmions}~(a), one has to integrate the
solid angle swept out by $\hat{\vect{M}}$:
\begin{equation}
\label{eq:windingnumber}
\mathcal{W} = \frac{1}{4 \pi} \int \hat{\vect M} \cdot
(\partial_x \hat{\vect M} \times \partial_y \hat{\vect M})\, dx\, dy.
\end{equation}
Comparing the winding number with the emergent fields [cf.\
Eq.~\eqref{eq:emergentBandEfield}] one observes that the emergent magnetic
field is of particular interest for the skyrmion lattice: The topology of
the skyrmion ensures the ``emergent magnetic flux'' per unit cell to be
quantized, as can be seen by integrating Eq.~\eqref{eq:emergentBfield}
over a magnetic unit cell (UC):
\begin{equation}
\int_{\text{UC}} \vect B^e \cdot d \vect \sigma = 4 \pi \hbar \, \mathcal{W},
\end{equation}
with $\mathcal{W}=-1$ for skyrmions in the chiral magnet
MnSi\cite{Muehlbauer2009, Neubauer2009, Yu2010} where the $z$ axis is along 
the external magnetic field direction. The topological Hall
effect due to the emergent magnetic field was one of the first
experimental confirmations that the magnetic textures observed in MnSi and
other B20 structures have a topological character and are actually
lattices of skyrmion tubes.\cite{Neubauer2009, Lee2009} Note that the
emergent fields~\eqref{eq:emergentBandEfield} have been derived for a
simple model, and therefore only a qualitative understanding of the
experimental signatures is possible at this level. To obtain a
quantitative agreement with experiments corrections due to non-adiabatic
processes\cite{Jiang2006, Schulz2012}, fluctuations of the magnetization
amplitude, modifications due to the band structure, etc. have to be taken
into account.

\subsection{Moving skyrmions and emergent electric field}
\label{ssec:EmergentElectricField}

As mentioned above, to observe the emergent electric field a
time-dependent magnetic texture is needed. One way to obtain a moving
magnetic texture is to exploit the spin-torque effect\cite{Slonczewski1996,
*Berger1996} by sending a sufficiently large spin-polarized
current through the sample which ``pushes'' the magnetic structure
forward.
Usually a spin polarized current is obtained by sending an electric
current through a sample with a finite magnetic moment, but there
are also other options to create spin currents like thermal or magnetic
field gradients or via the intrinsic spin Hall
effect.\cite{Dyakonov1971, *Hirsch1999, *Zhang2000, *Sinova2006, Murakami2003, *Sinova2004,
*Culcer2004}
The spin torque effect allows, for example, to move
ferromagnetic domain walls\cite{Grollier2003, *Tsoi2003} with the potentially
application of racetrack memories.\cite{Parkin2008} However, the threshold
current density, above which the domain walls get unpinned from disorder,
is very high, $j_c \sim \unit{10^{11}}{\ampere/\meter^2}$.

A key feature of the smooth skyrmion lattice in chiral magnets is the very
efficient coupling to electric currents. This is reflected in an ultra-low
threshold current density above which the skyrmion lattice starts moving.
For example, in MnSi,\cite{Jonietz2010, Schulz2012} the threshold current
density is five orders of magnitude smaller compared to the one for
traditional spin-torque effects, $j_c \sim \unit{10^{6}}{\ampere/\meter^2}$.

For current densities well above the threshold current density, a rigid
skyrmion lattice moves mainly along the current direction with a drift
velocity $\vect{v}_d$, and the relative speed between the spin current and
the magnetic structure is reduced.\cite{Schulz2012, PhdThesisEverschor,
Everschor2012, Iwasaki2013} For a rigid drifting magnetic texture with
$\hat{\vect{M}}(\vect{r}, t) = \hat{\vect{M}}(\vect{r} - \vect{v}_d t)$
the emergent electric and magnetic fields [Eq.~\eqref{eq:emergentBandEfield}]
are coupled since
$\partial_{t} \hat{\vect{M}} = -(\vect{v}_{d} \cdot \nabla) \hat{\vect{M}}$:
\begin{equation}
\vect E^e = -\vect v_d \times \vect B^e.
\label{eq:faradaylaw}
\end{equation}
Thus, one can identify the emergent electric field of the skyrmion lattice
by a reduction of the Hall signal,\cite{Zang2011, Schulz2012} because the
effective forces on the electrons due to the emergent electric field are
basically in opposite direction than the ones due to the emergent magnetic
field. Note that Eq.~\eqref{eq:faradaylaw} reflects Faraday's law of
induction, indicating that a change of the magnetic flux causes an
electric field. Another interesting aspect is that one can determine the
drift velocity $\vect{v}_d$ from the same Hall measurement by virtue of
Eq.~\eqref{eq:faradaylaw} since the emergent magnetic field of the skyrmion
lattice is quantized. For more details and experimental results see,
\textit{e.g.}, Ref.~\citenum{Schulz2012}.

\subsection{Spintronics with skyrmions}
\label{ssec:SkyrmionSpintronics}

So far we have discussed the Berry phase physics a conduction electron
experiences when traversing a spatially and temporally inhomogeneous,
smooth magnetic texture. Let us now switch the perspective and consider
the effects of the electrons on the magnetic texture. 
The magnetization dynamics can be modeled within the 
Landau-Lifshitz-Gilbert equation, an effective equation of motion 
for the slowly and smoothly varying magnetization direction of a 
magnetic texture with constant amplitude.\cite{Slonczewski1996, *Berger1996,
Zhang2004, *Duine2007, *Lucassen2009}
Here, we will not discuss the magnetization
dynamics in detail, but rather consider the current-induced forces on a
skyrmion texture from a qualitative point of view.
The equation of the forces (per skyrmion, per length, and per $\hbar$)
acting on the skyrmion lattice can be obtained from the 
Landau-Lifshitz-Gilbert equation using the Thiele method and treating 
pinning physics phenomenologically (see, \textit{e.g.}, 
Refs.~\citenum{Everschor2012, Schulz2012, PhdThesisEverschor}):
\begin{equation} 
 \vect{\mathcal D} (\tilde{\beta}{\vect v}_s-\tilde{\alpha} {\vect v}_d )
+ \vect G \times ( {\vect v}_s-{\vect v}_d ) 
+\vect F_{\mathrm{pin}} =0.
\label{drift1} 
\end{equation}
Here, $\vect{\mathcal D}$ is the dissipative 
tensor, 
$\tilde{\alpha}$ and $\tilde{\beta}$ are effective damping parameters,
$\vect v_s$ is an effective spin velocity, $\vect v_d$ is the drift 
velocity of the magnetic texture, $\vect G$ is the gyrocoupling vector, 
and $\vect F_{\mathrm{pin}}$ is the phenomenological pinning force.

\begin{figure}[tb]
\centering
\includegraphics[width=0.4\textwidth]{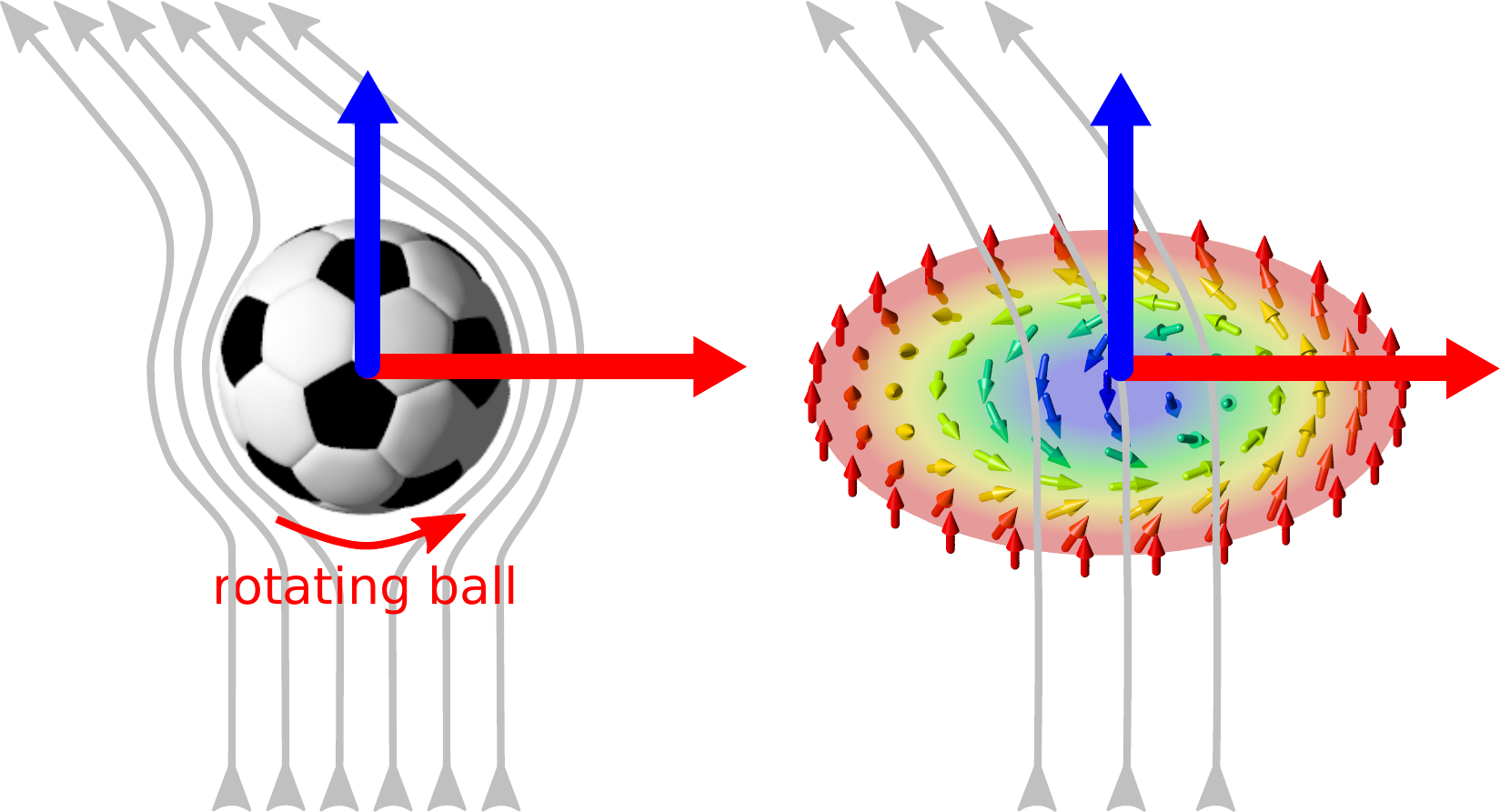}
\caption{%
Sketch of the current-induced forces acting on a skyrmion (right panel) in
comparison with the forces acting on a spinning ball in viscous air
(left panel).}
\label{fig:Magnuseffekt}
\end{figure}

Basically there are three different types of forces acting on the skyrmion
lattice when subjected to an electric current: two types of
current-induced forces 
(schematically shown in the right panel of Fig.~\ref{fig:Magnuseffekt}, 
for a non-moving skyrmion) and pinning forces. In the absence
of a current ($\vect v_s=0$), the skyrmion lattice is pinned by impurities and
the underlying atomic lattice, \textit{i.e.}, $\vect v_d=0$. Turning 
on the electric current, the current-induced forces are first fully 
compensated by the pinning force $\vect F_{\mathrm{pin}}$.\cite{FOOTNOTE3}
For larger current densities, the current-induced forces overcome the
pinning forces and induce a translational motion of the magnetic texture:
The first term of Eq.~\eqref{drift1} describes a dissipative or drag force 
draging the skyrmion lattice along the direction of the spin-current,
while the second term of Eq.~\eqref{drift1} describes a force on the 
magnetic texture, which is commonly referred to as a Magnus force. 
It can be understood (according to Newton's third law) as the 
counter-force on the magnetic texture opposite to the Lorentz force on 
the electrons due to the emergent electric fields. 
Hence, this Magnus force is related to Berry phase physics,
and occurs, in principle,
for any inhomogeneous magnetic texture. However, it is not surprising
that the Magnus force is also particularly interesting in skyrmion
lattices due to their topology. For the case of the skyrmion lattice, 
the gyrocoupling can be explicitly expressed in terms of the emergent 
magnetic field, $\vect G=\int_{UC} d^2 r M \vect B^e \sim \mathcal{W}$,
and is thus proportional to the winding number:\cite{Jonietz2010, Everschor2011}
Thus, for zero drift velocity the Magnus force is perpendicular to the 
direction of the spin current.

Note that the notion of a Magnus force is chosen in analogy to classical 
mechanics where, for example, the Magnus force acting on a spinning ball in 
viscous air leads to a bending of the ball's flight path 
(see Fig.~\ref{fig:Magnuseffekt}, left panel)
-- famous examples are ``banana kicks'' in soccer. In the rest-frame of
the ball, the origin of the Magnus force (depicted as the red thick 
arrow in the left panel of Fig.~\ref{fig:Magnuseffekt}) can be understood on a fairly 
simple level: For a non-spinning ball, the
air passes the ball on both sides equally fast, and thus there are no net
forces acting on the ball in perpendicular direction. However, for a
rotating ball the air is accelerated on one side and decelerated on the
other side so that the air is deflected. As the air experiences a force,
the ball experiences the opposite force -- the Magnus force -- according
to Newton's third law.

In the skyrmion case it is the Lorentz force induced by the emergent 
magnetic field that deflects the current and which in turn leads to a 
counter-effect -- the Magnus force -- on the magnetic texture.
Note, however, that this analogy is not complete, because, \textit{e.g.}, 
spin is not
a conserved quantity. Nevertheless, this picture helps to understand why
skyrmion lattices couple efficiently to electric currents. Current-induced
spin-torque effects occur only when the magnetization changes in space or
time. For spin structures like magnetic domain walls this is only a
nanoscopic region, while in the skyrmion lattice the smooth twist of the
local magnetization extends over macroscopic magnetic domains. In
addition, a weak coupling to the underlying atomic
lattice\cite{Muehlbauer2009} and a small pinning to
disorder\cite{Iwasaki2013} leads to the very low threshold current density
for the smooth skyrmion lattice.\cite{Jonietz2010, Schulz2012} This
excellent coupling to electric currents makes skyrmion systems very
interesting to study spin-torque effects as effects like Joule heating and
Oersted magnetic fields created by the current are less disturbing.
Furthermore, it allows for fundamental spin-torque experiments in bulk
materials, where surface effects that dominate in nanoscopic samples are
small.

With that knowledge about skyrmions, their excitations\cite{Petrova2011,
*Mochizuki2012, *Onose2012, *Mochizuki2013}, and how to
move,\cite{Schulz2012} rotate,\cite{Jonietz2010, Everschor2012} and even
merge them,\cite{Milde2013} skyrmions are not only a powerful laboratory
system to study fundamental physics, but may really become relevant for
technological applications.  Keeping skyrmion spintronics in mind, small
devices and thus small magnetic textures (where also momentum space Berry
phases are potentially relevant) are of course desired. Experimentally, it
has been shown that the size of the skyrmions can be controlled to some
extent by doping within the B20 structures.\cite{Muenzer2009, *Shibata2013}
On the atomic scale, skyrmions have been observed as the ground state of
an Fe film on Ir(111) substrates.\cite{Heinze2011} As a starting point for
magnetic storage technologies, the same group has recently achieved to
write and erase single skyrmions by surface techniques.\cite{Romming2013}
Furthermore, a potential generalization of racetrack memories using
skyrmions has been recently proprosed and investigated within
micromagnetic simulations.\cite{Fert2013} As those magnetic skyrmions are
of two-dimensional nature, it might also be interesting to
look for new, two-dimensional storage devices, where the intrinsic
properties of the skyrmions can be used even better.

To summarize, skyrmions
exhibit a lot of exicting physics and are tailored to study Berry-phase
and spin-torque effects.

\acknowledgments

We thank Achim Rosch and Christian Pfleiderer for continuous great support.
This work was supported by a fellowship within the postdoc program of the 
German Academic Exchange Service (DAAD). Financial support from the 
Deutsche Telekom-Stiftung (KE) is gratefully acknowledged.


%

\end{document}